\documentclass{article}
\usepackage{hyperref}
\pdfoutput=1
\usepackage{spconf,amsmath,graphicx} 

\usepackage{enumitem}
\setlist{nosep, leftmargin=14pt}

\usepackage{mwe} 
\usepackage{booktabs} 
\usepackage{multirow}
\usepackage{tabularx}
\usepackage{adjustbox}
\usepackage{array}
\usepackage{svg}
\usepackage{gensymb}
\usepackage{anyfontsize}
\usepackage{amssymb}
\usepackage{pifont}
\usepackage{algorithm,algorithmicx}
\usepackage[noend]{algpseudocode}

\usepackage{makecell}

\title{White Light Specular Reflection Data Augmentation for \\Deep Learning Polyp Detection}
\name{
  Jose Angel Nu\~nez$^{1}$  
  Fabian Vazquez$^{1}$ 
  Diego Adame$^{1}$  
  Xiaoyan Fu$^{2}$ 
  Pengfei Gu$^{1}$  
  Bin Fu$^{1}$
}

\address{
  $^{1}$Department of Computer Science, University of Texas Rio Grande Valley, Edinburg, TX, USA \\
  $^{2}$The Second Affiliated Hospital of Fujian University of Traditional Chinese Medicine,\\
  Fuzhou, Fujian, China
}
\begin{document}
%

\maketitle
\begin{abstract}
Colorectal cancer is one of the deadliest cancers today, but it can be prevented through early detection of malignant polyps in the colon, primarily via colonoscopies.
While this method has saved many lives, human error remains a significant challenge, as missing a polyp could have fatal consequences for the patient.
Deep learning (DL) polyp detectors offer a promising solution.
However, existing DL polyp detectors often mistake white light reflections from the endoscope for polyps, which can lead to false positives.
To address this challenge, in this paper, we propose a novel data augmentation approach that artificially adds more white light reflections to create harder training scenarios. 
Specifically, we first generate a bank of artificial lights using the training dataset. 
Then we find the regions of the training images that we should not add these artificial lights on.
Finally, we propose a sliding window method to add the artificial light to the areas that fit of the training images, resulting in augmented images.  
By providing the model with more opportunities to make mistakes, we hypothesize that it will also have more chances to learn from those mistakes, ultimately improving its performance in polyp detection.
Experimental results demonstrate the effectiveness of our new data augmentation method.

\end{abstract}
%

%
\section{Introduction}
\label{sec:introduction}
Colorectal cancer (CRC) is a significant global health issue, which is the third most common cancer worldwide in 2021 and the second leading cause of cancer-related deaths~\cite{siegel2023colorectal,WHO2021,zhang2024ihcsurv,gu2024boosting,wang2024path}.
The key to preventing CRC is the early detection and removal of adenomatous polyps.
Colonoscopies are known as the gold standard for detecting polyps~\cite{zauber2015impact, huck2016colonic,pacal2021robust}, but the procedure's effectiveness depends heavily on the experience of the endoscopist.
Even experienced endoscopists miss about 12\% of large adenoma polyps (at least 1 cm)~\cite{bernal2021}, and missed polyp rates can vary between 15\% to 35\%, depending on the size of the polyp~\cite{lalinia2024colorectal}.

Deep learning (DL) object detectors are capable of locating and classifying objects of interest~\cite{liang2020intracker}. 
There are two types of DL object detection algorithms:
Single-stage detectors, such as YOLO (You Only Look Once)~\cite{redmon2016you} and its variants and two-stage detectors. such as Faster R-CNN~\cite{ren2016faster} and its variants.
Single-stage detectors are useful for real-time
applications like polyp detection during colonoscopy due to their speed and efficiency.  
These detectors perform object localization and classification in one step, allowing them to process images faster. 
However, this could come at the cost of accuracy, especially when detecting small, flat, or occluded polyps that may not be well-defined. 
Single-stage detectors can also suffer from higher false positive rates in challenging conditions, such as reflections or poor lighting, which are common in medical imaging \cite{pacal2021robust}.
Two-stage detectors offer higher accuracy by separating region proposal from classification.
This makes them known for more complex tasks where fine-grained detection is necessary. 
However, two-stage detectors are slower due to their sequential processing, making them harder to use  for real-time applications. 
They are also computationally more expensive, which limits their practicality in scenarios that require immediate feedback during procedures. 
Although advancements like spatial pyramid pooling have improved their speed, they still lag behind single-stage models in real-time performance \cite{he2015spatial}.
\begin{figure}[t]
    \centering
    \includegraphics[width=0.42\textwidth]{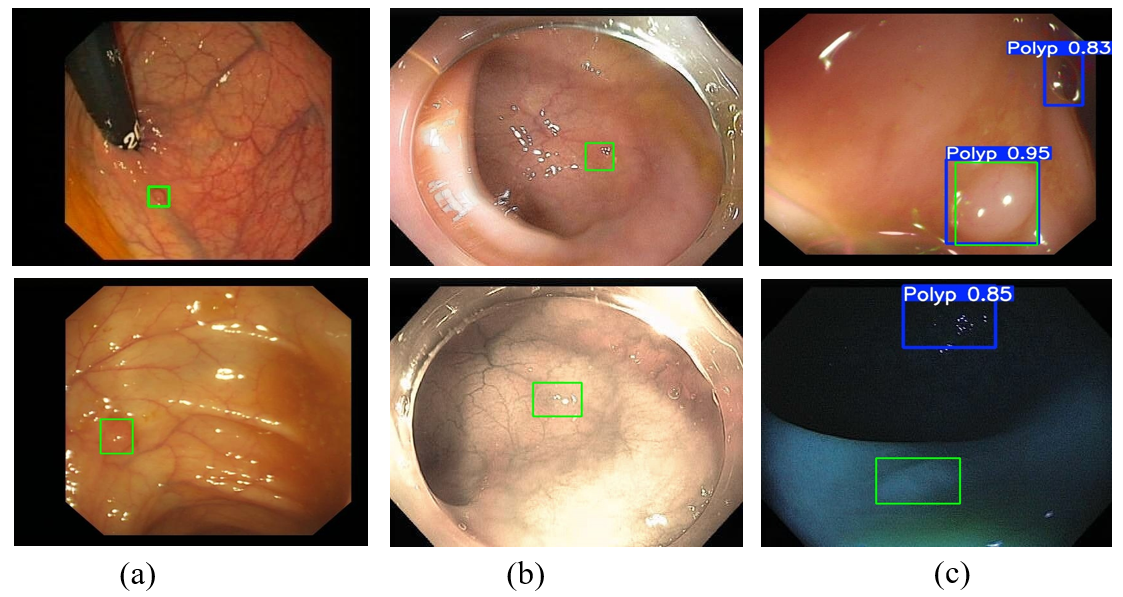} 
    \caption{Visual examples highlighting challenges in detecting polyps: (a) small polyps, (b) flat polyps, and (c) white light specular reflections causing confusion with polyps.}
    \label{fig:challenges}
\end{figure}
Leveraging DL detectors for polyps detection may aid endoscopists in reducing the missed polyp rate. 
Many researchers have explored variations of YOLO for polyp detection.
ResYOLO \cite{zhang2018polyp} incorporated residual connections and 
efficient convolution operators for polyp detection, achieving a precision of 88.6\% and a recall of 71.6\%.
\cite{nogueira2022real} utilized YOLOv3~\cite{farhadi2018yolov3} and an object tracking algorithm to reduce false positives by filtering out bounding boxes.
Although effective, they reported lower predictive performance on flat polyps.
\cite{pacal2021robust} proposed a robust real-time DL polyp detection system, using data augmentations, such as noise invocation, flipping, rotation, brightness, and contrast adjustment.
\cite{ghose2024improved} fine-tuned YOLOv5 variants for improving the detection of smaller polyps.

Many problems exist when attempting to detect polyps, such as a polyp being small or flat. 
Another common challenge is that polyp detectors tend to confuse the white light specular reflections produced by the endoscope with polyps (see Fig.~\ref{fig:challenges}). 
However, this challenge is a less explored topic. 
To address this challenge, in this paper, we propose a new data augmentation called White Light Specular Reflection (WLSR), for DL polyp detection.
The purpose of this data augmentation is to generate more specular reflections and add them to the training images to make training harder. 
Specifically, we first generate a bank of artificial lights using the training dataset. 
Then we find the regions of the training images that we should not add these artificial lights on.
Finally, we propose a sliding window method to add the artificial light to the areas that fit of the training images, resulting in augmented images.  
By providing the model with more opportunities to make mistakes, we hypothesize that it will also have more chances to learn from those mistakes, ultimately improving its performance in polyp detection.

Our contributions are as follows: (1) We develop a new data augmentation for DL polyp detection. 
(2) We demonstrate the effectiveness of our new method on the Harvard Dataverse dataset~\cite{DVN/FCBUOR_2021}.
\begin{figure}[t]
    \centering
    \includegraphics[width=\columnwidth]{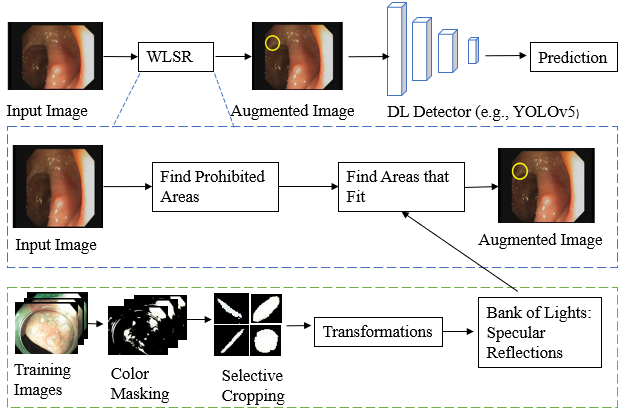} 
    \caption{The overview of our WLSR data augmentation for DL polyp detection.}
    \label{fig:pipeline}
\end{figure}

\section{Methodology}
\label{sec:pagestyle}
Fig.~\ref{fig:pipeline} shows the pipeline of our new data augmentation method: First, we generate a bank of artificial lights using the training dataset (Section~\ref{ssec:bank}). Then for input training images, we identify the regions in which we cannot place these lights (Section~\ref{ssec:prohibited}).
Next, we place these artificial lights to the areas that fit to form the augmented images. 
Finally, the augmented training images are used to train the DL detector (e.g., YOLOv5~\cite{ghose2024improved}) for polyp detection.

\subsection{Generating Bank of Lights}
\label{ssec:bank}
Simply placing white circles or ovals may not be beneficial for training since we need to mimic the real thing. 
Therefore, a crucial question is: How can we generate the artificial lights that look like real white light specular reflections that occur during colonoscopies?

To address this challenge, we propose to generate a bank of lights, consisting of the following steps:
(1) We identify regions of the real white light specular reflections of the training images.
(2) We generate various shades of white light on these regions by adjusting the intensity levels of the regions and turn regions that are not real white light specular reflections to black. 
Different real white light specular reflections regions have different grayscale values (from dark gray to bright white).
This can simulate a more realistic light effect with intensity variations.
(3) We selectively crop the regions of the white lights 
to ensure that we have different sizes and ample variations of white lights.  
Overall, 300 regions of generated white lights are cropped.
(4) Lastly, 11 transformations are applied to these 300 regions of white lights, forming the bank of lights.
The transformations are performed in 3 rounds. 
First round, the 300 original regions undergo these 4 transformations: 1) flipping horizontal, 2) flipping vertical, 3) random scaling (scaling between 0.8 to 1.2), and 4) random rotation (rotation range: -30 to 30 degrees).  
Each of the 4 transformations listed above transforms the original 300 regions: thus, 4 x 300 = 1,200 new images. 
In the second round, the images flipped vertically and horizontally get transformed by random scaling and random rotation to create more lights. 
This results in 4 x 300= 1,200 new images. 
In the third round, random scaling is performed on 3 different sets of images that underwent the following transformations: random rotation on the original images, random rotation on the image set flipped vertically, and random rotation on the image set flipped horizontally.
This results in 3 x 300 = 900 new images.
Therefore, in total, we have 300 original images + 1,200 from first round + 1,200 from second round + 900 from third round = 3,600. 

\subsection{Finding Prohibited Areas}
\label{ssec:prohibited}
There are three situations we have to consider when placing our artificial lights in the training images: 
(1) It must not overlap current lights because it could deform them and create the shapes of 
lights that don’t exist during colonoscopies.
(2) It must not overlap with any polyps to avoid possible occlusions of polyps.
(3) It must not be placed on the black borders considering that the training images in the dataset contain black borders of different sizes and shapes.

To satisfy the above conditions, we convert the prohibited areas to a color that does not appear during colonoscopy procedures;in this case, orange was chosen. 
We identify the prohibited areas outlined above using the following methods: 
(1) We generate bounding boxes for the light areas in the images by first creating segmented images using a binary mask that highlights light regions based on a threshold range for shades of white. 
These segmented images are then processed to find contours representing the light areas, and bounding boxes are computed around these contours. 
Finally, we refine the bounding boxes using non-maximum suppression to eliminate overlaps and redundancies.
We mark the final bounding box areas in orange. 
(2) We turn polyp areas orange by using their bounding boxes from our training dataset annotations.
(3) We convert the black borders of images to orange by first creating a binary mask that selects all black pixels in the image.
It then iterates over each pixel using nested loops, checking two conditions: whether the pixel is black (as indicated by the mask) and whether it is within the outer 20\% margins of the image dimensions (top, bottom, left, or right).
If both conditions are true, the pixel's color in the original image is changed to orange. 

\subsection{Finding Areas that Fit}
\label{ssec:subsubhead}

Once we identify the prohibited areas (marked by orange pixels), we proceed to locate the areas that fit.

We accomplish this via a sliding window approach. 
We first generate a random artificial light from the bank of lights and then starting from the top left corner it moves in sliding window fashion. 
From the left to right until it reaches the end of the image and then it steps down and follows the same pattern from left to right until all areas of an image have been covered.
The step size to move horizontally is the width of the cropped light image plus one.
The step size to move vertically is the height of the cropped light image plus one.

If at any point during the sliding window process there is no orange pixel in the whole area that the cropped light image is covering, then the top left coordinate of that location gets stored in an array.
The array of top left coordinates provides us with all the areas that fit.
Lastly, from that array it randomly picks one top left coordinate where we are going to place the artificial light.
We decided to do only one additional artificial light to see if it results in at least a minimal improvement to the model.
In the future, we plan to test it with more than one artificial light added to images. 

\begin{figure}[ht]
    \centering
    \includegraphics[width=\columnwidth]{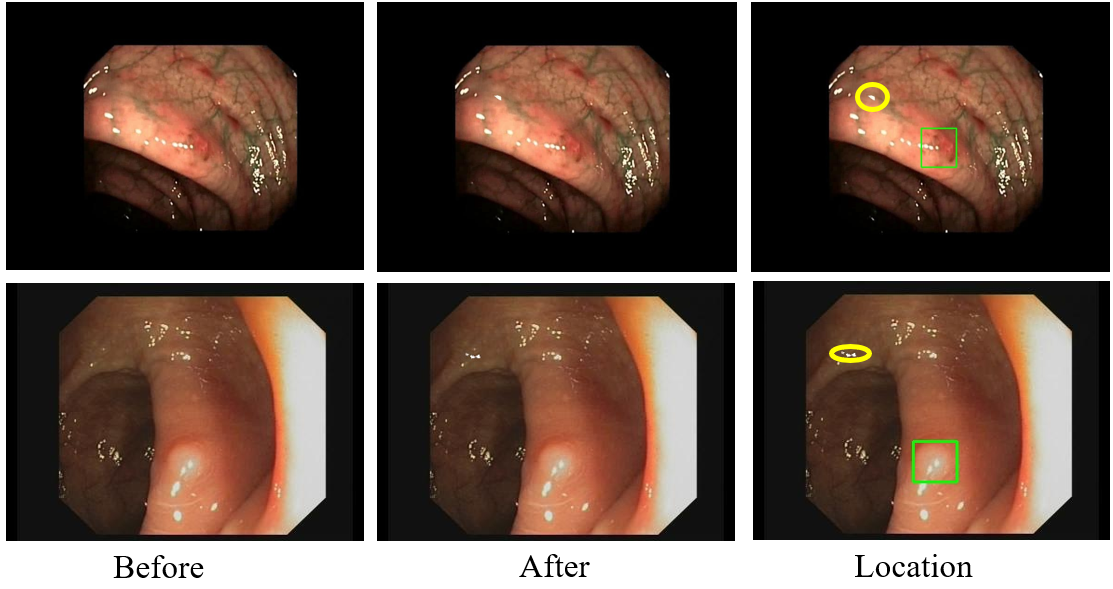} 
    \caption{WLSR data augmentation results. Yellow circles indicate the new lights generated by our method, while green boxes highlight the real polyps.}
    \label{fig:da-results}
\end{figure}

If no areas fit, the algorithm tries again n number of times with different artificial lights from Bank of Lights.
After n tries, that image and its respective annotation are not used for training.
Only the non-black portions are pasted onto the image.
Fig.~\ref{fig:da-results} shows two results of the WLSR data augmentation, along with an extra image that circles where the new light is located.

\begin{table*}[t]
\centering
\begin{tabular}{lcccccccc}
\hline
Method                                   & Precision (\%)  & Recall (\%)   & mAP50 (\%)    & mAP50-95 (\%) \\ \hline
Training w/ original images     & \textbf{87.6}            & 63.6          & 77.5          & 54.5     \\ 
Training w/  100\% WLSR Data                        & 82.8            & 67.2          & 78.4          & 53.9    \\\hline
Training w/  10\% WLSR Data                         & 84.3            & 65.0          & 78.2          & 53.8     \\ 
Training w/  20\% WLSR Data                        & 83.1            & 71.5          & 80.7          & 54.8    \\ 
Training w/  50\% WLSR Data                         & 84.6            & 67.1          & 79.1          & 54.8    \\ \hline
Training w/  20\% WLSR Data  + Other DAs   & 86.5            & \textbf{71.7} & \textbf{81.5} & \textbf{55.9}     \\ \hline
\end{tabular}
\caption{Results of different methods. The best performance is in \textbf{bold}.}~\label{table1}
\end{table*}

\section{Experiments and Results}
\label{sec:majhead}

\textbf{Dataset.} 
The data used for this research comes from the Harvard Dataverse~\cite{DVN/FCBUOR_2021}. 
This dataset comprises polyp images sourced from MICCAI 2017, CVC Colon DB, GLRC, KUMC, and the University of Kansas Medical Center, each accompanied by XML annotations. 
The XML annotations describe important details such as coordinates of where the polyp is located in an image. 
This data contains three main folders: train 2019, validation 2019, and test 2019. 
The XML annotations got converted to YOLO annotation format. 
Some images were not used because they either did not have bounding box coordinates for their respective image or did not meet a certain condition during the WLSR data augmentation pipeline. 
In the end, our dataset consisted of 26,942 training images, 4,214 validation images, and 4,685 test images.

\textbf{Implementation Details.} 
We trained all experiments using two different computers, one computer with an NVIDIA GeForce GTX 1080 GPU, and the second computer was equiped with an NVIDIA GeForce RTX 2080Ti GPU. 
All the models that were trained utilized the pre-trained COCO model as a starting point.
This was done to speed up training, and because the data size for polyps is small, it is highly recommended to start training on a pre-trained model.
Default training parameters, test parameters, and hyperparameters were used in these experiments.
We kept them consistent across all experiments to clearly see if models with WLSR data augmentation included offered an increase in performance metrics.
The default data augmentations were also disabled
for all models except the last one (Training w/20\% WLSR Data + Other DAs).
Default parameters were, image size=640, batch size=16, confidence threshold=0.25, IoU threshold=0.45, optimizer= stochastic gradient descent, initial learning rate = 0.01, final learning rate=0.2, momentum = 0.937, weight decay (L2 regularization)=0.0005, warm up epochs =3, warm up momentum =0.8, warm up bias learning rate=0.1, box loss gain=0.05, class loss gain=0.5, class loss positive weight=1, objectness loss gain=1, objectness loss positive weight=1, IoU threshold for matching=0.2, and anchor matching threshold=4. 

All models were training for 30 epochs.
Each model got trained 3 different times with the seed values of 1, 2, and 3, then an average of the results was recorded. 

\textbf{Experimental Results with 100\% Replacement.} To evaluate our approach, we replace 100\% of the original training data by performing the WLSR data augmentation to it and then training with those images.
As shown in Table~\ref{table1}, one can see that compared with the model that is trained with the original training data, there is 0.90\% improved in mAP50.
This demonstrates the effectiveness of our WLSR data augmentation method.

\textbf{Experimental Results on Exploring the Limits.} 
We conduct experiments to investigate the limits of the WLSR augmentation by training the model with different percentages of images that get augmented by WLSR.
Specifically, we train the model with 10\%, 20\%, and 50\% replacement. 
From Table~\ref{table1}, we can notice that training with 20\% replacement produces the best performance, resulting in 3.2\% improvement in mAP50.
This confirms our hypothesize that by providing the model with more opportunities to make mistakes, it will also have more chances to
learn from those mistakes, ultimately improving its performance in polyp detection.

\textbf{Effects With Other Data Augmentations.} 
We conduct experiments to investigate the effects with other data augmentations.
Specifically, we train the model with 20\% replacement and other data augmentations: hsv\_s=0.1(saturation augmentation + or - 10\%), hsv\_v= 0.1(value augmentation + or - 10\%), degrees=5(rotation augmentation + or - 5 degrees), and fliplr=0.5(flip left or right with 50\% chance). 
Table~\ref{table1} shows that when combining our WLSR with other data augmentation methods, there is a 0.8\% further improvement in mAP50.
This suggests that our new WLSR data augmentation method could be used as a plug-in approach.

\begin{figure}[ht]
    \centering
    \includegraphics[width=\columnwidth]{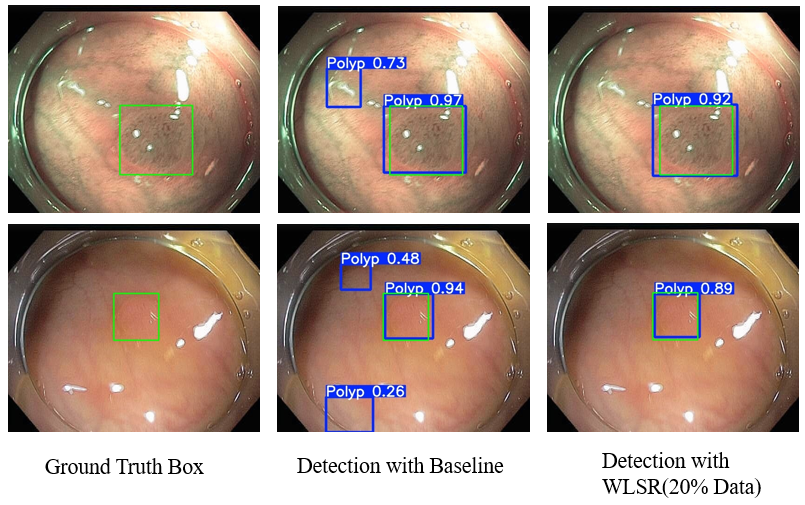} 
    \caption{Visual results demonstrating the effectiveness of our method.}
    \label{fig:fina-vis-results}
\end{figure}

\textbf{Visual Results.} 
Fig.~\ref{fig:fina-vis-results} shows two examples of polyp detection when training the model with original training images and with 20\% replacement.  
From the visual results, we can observe that the model trained with original training images may confuse a specular reflection with a polyp, while the model with WLSR did not make that mistake. 

\section{Conclusion}
\label{sec:page}
In this paper, we presented a new data augmentation named WLSR for DL polyp detection.
Our method is capable of adding more specular reflections to training images to make training harder and hence to give the model more chances to learn new features that would help it distinguish a polyp from a specular reflection. Experimental results on one public dataset demonstrate the effectiveness of our new data augmentation method.

\section{Compliance with ethical standards}
\label{sec:ethics}
This research study was conducted retrospectively using human subject data made available in open access by one publicly available dataset~\cite{DVN/FCBUOR_2021}. Ethical approval was not required as confirmed by the licenses attached with the open access datasets.

\section{Acknowledgements}
\label{sec:acknowledgements}
This work was supported in part by the Graduate Assistance in Areas of National Need (GAANN) grant.
\bibliographystyle{IEEEbib_abbrev}
\bibliography{refs}

\end{document}